# High Numerical Aperture and Broadband Achromatic Flat Lens

Jingen Lin [1]†, Jinbei Chen [1]†, Jianchao Zhang [1,3], Haowen Liang [1]*, Juntao Li [1,2]*, Xue-Hua Wang [1,2]*

[1] State Key Laboratory of Optoelectronic Materials and Technologies, School of Physics, Sun Yat-Sen University, Guangzhou 510275, China.

[2] Quantum Science Center of Guangdong-Hong Kong-Macao Greater Bay Area (Guangdong), Shenzhen, China.

[3] Hisense Laser Display Co.,Ltd, 399 Songling Road, Qingdao, Shandong, China.

† These authors contributed equally to this work.

*Corresponding authors. lianghw26@mail.sysu.edu.cn (H.L.); lijt3@mail.sysu.edu.cn (J.Li); wangxueh@mail.sysu.edu.cn (X.H.W.)

**Abstract:** Flat lenses have shown promising applications in miniaturized and ultracompact lightweight optical systems. However, it has been a great challenge in simultaneously achieving broadband achromatism and high numerical aperture. Here, we demonstrate that this long-term dilemma can be broken through by the zone division multiplex of the meta-atoms on a composite substrate possessing stepwise optical thickness. The aperture size can be freely expanded by increasing the optical thickness difference between the central and marginal zones of the substrate, free from achromatic bandwidth. The achromatic flat lens with both 0.9 numerical aperture and bandwidth of 650-1000 nm is experimentally achieved. A microscopic imaging with 1.1 µm resolution has also demonstrated. These unprecedented performances mark a substantial step toward practical applications of the flat lenses.

# INTRODUCTION

Miniaturized, ultracompact, and lightweight optical systems[1-3], such as chip-scaled imagers[4], portable displays[5], and wearable optical devices[6], are increasingly desired in modern demands. The crucial elements in these systems are the optical lenses that focus light achromatically. Although achromatic lenses were developed centuries ago, they necessitate the assembly of multiple components to control light dispersion. This results in bulky and intricate systems that fail to meet the demands of miniaturization, ultra-compactness and lightweight. Therefore, single-lens systems that simultaneously deliver multiple high optical performances, such as wide achromatic bandwidth, high numerical aperture (NA), and a slim form-factor, emerge as promising candidates to satisfy contemporary demands. Flat lenses, notable for their compactness and ease of integration, stand out as ideal candidates for such single-lens systems[7-10]. Unlike traditional optical lenses, flat lenses excel in delivering precise and efficient phase control[11-15]. This capability allows for the manipulation of transmitted or reflected light, enabling the achievement of specific functionalities. In many imaging situations, the incident light is chromatic, requiring achromatic focusing by flat lenses to ensure clear imaging[16, 17]. However, a lens designed based on the phase equation for a specific operational wavelength will encounter focal spot deviations when the actual operating wavelength differs from the intended design wavelength. This issue, known as chromatic aberration, tends to be more pronounced in flat lenses than in traditional optical lenses.

Achromatic metalenses represent a category of achromatic flat lenses constructed from sub-wavelength meta-atoms[18-31]. These meta-atoms display different phase responses across various wavelengths, enabling the correction of chromatic aberration through the adjustment of wavelength-dependent phase dispersion. By utilizing the same diffraction order, the metalens focuses all wavelengths at a single point, effectively countering the issue of chromatic aberration. It is important to note that achromatic metalenses have faced limitations regarding their NA, operating bandwidth, and size due to the constraints in the phase dispersion capabilities of meta-atoms[23, 32-34]. This limitation restricts the use of an achromatic metalens in achieving both high NA and broad bandwidth simultaneously within a specific size constraint. Many previous studies have focused extensively on expanding the phase dispersion capabilities of the meta-atoms[19, 21, 22, 26, 35]. However, this improvement is very limited, and inevitably complicates both the development of a meta-atoms library and the fabrication process. An alternative type of achromatic flat lens is the achromatic multi-level diffractive lens (MDL)[36, 37]. For instance, large-scale achromatic MDLs have been designed and demonstrated to operate over a wide wavelength range, with lens sizes up to 1 cm and an NA of 0.1[37]. MDLs are capable of focusing different discrete wavelengths at the same focal point by utilizing various diffraction orders. This distinctive feature gives MDLs an advantage in creating large-size achromatic flat lenses. However, this comes with significant challenges when trying to achieve a high NA, including the high complexity of fabrication and a notable decrease in efficiency. By using both types of these flat lenses, even when leveraging multilayer or hybrid designs built upon single-layer configurations[12, 21, 38-40], creating a broadband achromatic flat lens that simultaneously meets the criteria for high NA and imaging resolution beyond 2 μm in both the visible and near-infrared (NIR) ranges continues to be a significant challenge to date (see also Table 1).

In this work, we present the zone division multiplexing strategy of the meta-atoms on a composite substrate partitioned by stepwise optical thickness to break through the long-term dilemma between high NA and broad achromatic bandwidth, as shown in Fig. 1. In this strategy, the phase dispersion required by the achromatic flat lens are jointly compensated by the meta-atoms and stepwise optical thickness of the substrate. The diameter of the zone division multiplexing flat lens can be on-demand extended by increasing the optical thickness difference between the central and marginal zones of the composite substrate, free from the limit of the achromatic bandwidth, which facilitates the creation of a high-performance achromatic flat lens. Based upon this ideal, we have experimentally fabricated an achromatic flat lens with an NA of 0.9 and a radius of 20.1 µm with using a maximum meta-atoms phase dispersion compensation of only 2.2π. Its excellent achromatic focusing capability spans the entire first NIR window, covering wavelengths from 650 nm to 1000 nm. For the conventional achromatic metalens, such high performance requires an unachievable phase dispersion compensation of up to 13.6π, while the phase dispersion compensation of 2.2π only create the achromatic flat lens with the NA of 0.56 and the radius of 6.6 µm. Meanwhile, the NA of the fabricated flat lens is significantly higher than that of the achromatic MDLs. Furthermore, the broadband microscopic imaging with high resolution of 1.1 µm has achieved by using the zone division multiplexing achromatic flat lens with the NA of 0.7 and radius of 30.0 µm. In contrast, both the conventional achromatic metalenses and MDLs face challenges in achieving imaging resolutions finer than 2 µm. Our approach introduces a new avenue for surpassing the existing limitations and challenges associated with creating high-performance achromatic flat lenses that simultaneously meet desired NA and size specifications.

## PRINCIPLE OF DESIGN

We investigate the flat lens constructed by the meta-atoms on a composite substrate with varying optical thickness, which satisfy the achromatic conditions of the spherical aberration elimination[18, 25, 41]. Assuming the radius of the flat lens is $R$, we choose the phase of meta-atoms at $R$ as zero within the achromatic range, then the phase profile of the achromatic flat lens $\varphi_{lens}(r,\omega)$ satisfies the following condition:

$$\varphi_{lens}(r,\omega) = \varphi_{meta}(r,\omega) + \frac{\omega}{c}L(r) = \frac{\omega}{c}\left[\sqrt{R^2+f^2} - \sqrt{r^2+f^2} + L(R)\right] \quad (1)$$

where $\varphi_{meta}(r,\omega)$ is the phase profile contributed by the meta-atoms at the radial position $r$; $\omega$ is the frequency of the light; $L(r) = \bar{n}(r)l$ is the optical thickness of the composite substrate at $r$; $\bar{n}(r)$ and $l$ are the average refractive index and the thickness of the composite substrate, respectively; $f$ is the designed focal length. It is noted that the optical thickness of the substrate imposes a very large initial phase into the actual phase of the flat lens. To eliminate this effect in the following analysis, we only consider the relative phase of the flat lens $\varphi_{lens}(r,\omega) - \varphi_{lens}(R,\omega)$, i.e., the output phase of the flat lens induced by the meta-atoms and the optical thickness between its radius at $r$ and $R$.

We denote the maximum phase dispersion that can be compensated by the meta-atom library as $\Delta\Phi_{meta}$, and the value of the phase dispersion is positive[23]. The required phase dispersion of the meta-atoms from the phase profile suggested in Eq. (1) is then subject to the following restrictions:

$$0 \leq \Delta\varphi_{meta} = \frac{\Delta\omega}{c}\left[\sqrt{R^2+f^2} - \sqrt{r^2+f^2} - \Delta L(r)\right] \leq \Delta\Phi_{meta} \qquad (2)$$

Where $\Delta\varphi_{meta}$ is the required phase dispersion of the meta-atoms, $\Delta\omega$ is the bandwidth of incident light, $\Delta L(r) = L(r) - L(R)$ is the optical thickness difference of the composite substrate between its radius at $r$ and $R$. Noticing from the right side of Eq. (1) that the maximum phase dispersion occurs at the center of the achromatic flat lens ($r = 0$), the radius $R$ of the achromatic flat lens can be obtained from Eq. (2):

$$\sqrt{R^2+f^2} = L_{meta} + \Delta L(0) + f, \qquad (3)$$

where $L_{meta} = c\Delta\Phi_{meta}/\Delta\omega$ is so-called meta-featured size, $\Delta L(0)$ the optical thickness difference between the center and edge of the flat lens. Evidently, when $\Delta L(0) = 0$ corresponding to the conventional metalens, the diameter of the flat lens is determined only by the meta-featured size $L_{meta}$ and limited by the achromatic bandwidth. Eq. (3) shows that the optical thickness difference $\Delta L(0)$ of the composite substrate provide a new degree of freedom to expand on-demand the radius of the achromatic flat lens without compromising bandwidth $\Delta\omega$ for a given $\Delta\Phi_{meta}$, which is a significant breakthrough to the limitation imposing on conventional achromatic metalenses.

We begin with taking a conventional achromatic metalens as an example, where $L(r) = $

$L(R)$ is fixed as a constant[23]. In this case, as shown in the left side of Eq. (2), the required phase dispersion decreases with increasing $r$ until reaching $R$, e.g. $R = r_1$, where it achieves the maximum radius of the conventional metalens. Taking into account the practical fabrication constraints in our actual foundry, we use the meta-atom library with the maximum achievable phase dispersion of $\Delta\Phi_{meta} = 2.2\pi$. We then use this library to design an achromatic flat lens with a focal length $f$ of 9.7 μm, and a bandwidth from 650 nm to 1000 nm. The corresponding relative phase of the flat lens $\varphi_{lens}(r,\omega) - \varphi_{lens}(R,\omega)$ is illustrated in Fig. 2A, and the maximum radius of the conventional metalens is limited to $r_1 = 6.6$ μm, which corresponds to an NA of 0.56 (the grey region in Fig. 2A). As shown in Fig. 2A, when $r > r_1$, $\Delta\varphi_{meta} < 0$, which requires the meta-atoms to compensate the negative phase dispersion and is beyond the compensation capacity of the meta-atom library. If we expand the size of the achromatic metalens to a radius $R$ of 20.1 μm, corresponding to an NA of 0.9, then the required maximum phase dispersion in the meta-atom library would need to reach $\Delta\Phi_{meta} = 13.6\pi$. For a conventional achromatic metalens in practice, such a design is extremely difficult to attain due to the excessively high phase dispersion exhibited by the meta-atoms, rendering experimental realization infeasible. This limitation constrains the development of conventional metalenses that simultaneously achieve high NA, broad bandwidth, and large size. This is a significant challenge in both the development of a meta-atom library and the actual fabrication process.

The key of extending the lens radius is to eliminate the negative phase dispersion. We then raise the idea of zone division multiplexing of the meta-atoms by dividing the flat lens into $N$ zones according to the reduce of the optical thickness $L(r)$ along the radial direction. Considering the discrete meta-atoms and the convenience of the actual fabrication difficulty of the composite substrate, the discrete strategy is adopted in this work. In this way, the working radius of the designed flat lens $r_N$ is equal to the radius $R$ and it is significantly larger than the maximum radius $r_1$ of a conventional metalens constructed by the same meta-atom library. Therefore, such strategy opens up the way that breaks through the limit on the lens radius incurred by the achromatic lens bandwidth, and can simultaneously achieve high NA and broadband by increasing the optical thickness difference of the composite substrate.

We then demonstrate that the above design can be achieved by an achromatic flat lens which multiplexes the same meta-atom library with a maximum phase dispersion of $\Delta\Phi_{meta} = 2.2\pi$. The composite substrate is constructed according to the zone division of the phase profile of the meta-atoms. The details of the design refer to Section S2 in the Supplementary Materials. According to the design, the phase profile contributed by the meta-atoms (as shown the solid lines of Fig. 2B) is discontinuous; therefore, the relative phase $\Delta\omega\Delta L(r)/c$ imposed by the optical thickness difference of the stepwise substrate should also be discontinuous (as illustrated by the dashed lines in Fig. 2B) so that the relative phase of the flat lens $\varphi_{lens}(r,\omega) - \varphi_{lens}(R,\omega)$ is continuous, as depicted in Fig. 2C. Fig. 2C suggests that this achromatic flat lens successfully enhances the maximum NA from 0.56 to 0.9 and expands the radius from 6.6 μm to 20.1 μm by multiplexing the same meta-atoms library, surpassing the capabilities of a conventional achromatic metalens. This improvement is attained without compromising the achromatic wavelength range from 650 nm to 1000 nm. Therefore, we provide an effective way to achieve an achromatic flat lens with both a higher NA and larger size by incorporating additional step zones or using meta-atoms capable of a larger phase dispersion. This development successfully breaks through the long-term dilemma in the achromatic flat lens that cannot simultaneously have high numerical aperture and broad bandwidth.

## SIMULATION, FABRICATION, AND CHARACTERIZATION

The meta-atoms within the metalens consist of crystalline silicon (c-Si) nanopillars situated on a $SiO_2$ substrate, arranged with a periodicity of 300 nm. To safeguard the structure, a layer of hydrogen silsesquioxane (HSQ) is applied and cured atop the nanopillars. The height of the c-Si nanostructures is established at 500 nm. The choice to use c-Si is driven by its high refractive index[42], which allows the meta-atoms to demonstrate significant phase dispersion. To enhance phase dispersion further and ensure polarization insensitivity, five types of symmetrical meta-atoms are utilized, as shown in Fig. 3A. These meta-atoms achieve full $2\pi$ phase coverage and a maximum phase dispersion of $2.2\pi$, as detailed in Fig. S3 of the Supplementary Materials. Following this, the meta-atoms are organized according to the phase profile depicted in Fig. 2B and then are seated onto the composite substrate according to the zone division. This configuration yields to the achromatic flat lens illustrated in Fig. 1, which boasts an NA of 0.9 and a radius of 20.1 μm.

To assess the achromatic performance of the flat lens, finite difference time-domain (FDTD) simulations are performed using the commercial software FDTD Solutions (Lumerical Inc.), employing a horizontally linearly polarized (x-polarized) incident beam. The numerical simulation results are depicted in Fig. 3B to E. Figures 3B to C display the point spread functions (PSFs) of the focal spot, covering a wavelength range from 650 nm to 1000 nm. It has been observed that the input linearly polarized beam is focused by the lens into a flat elliptical spot. This phenomenon is commonly associated with high NA lenses and originates from the vector characteristics of focused spot. In this case, we use Richards-Wolf vector diffraction integration method (VDIM) to calculate and characterize the diffraction limit focused spot of the ideal lens with high NA, rather than the Airy disk mode [43, 44]. As depicted in Fig. 3D, the full width at half maximum (FWHM) and Strehl ratios (SR) of the achromatic flat lens meet the diffraction limit across the full spectrum of operational wavelengths. The red dots in Fig. 3E highlight the focal length and depth of focus (DOF) of the achromatic flat lens. For comparison, the blue dots in Fig. 3E depict the outcomes using a conventional achromatic metalens of the same size and NA, which relies exclusively on the phase profile illustrated in Fig. 2A. Evidently, our specially designed achromatic flat lens surpasses conventional metalens designs in correcting chromatic aberration. This clearly demonstrates that our flat lens achieves nearly ideal chromatic aberration correction and diffraction-limited focusing within its intended achromatic wavelength range, rendering it highly suitable for high-performance imaging systems.

Experimental fabrication and measurements were conducted to validate the performance of the achromatic flat lens, which has an NA of 0.9 and a radius of 20.1 µm. The fabrication process details are provided in the Materials and Methods section and Fig. S4 in the Supplementary Materials. The resulting structure is displayed in Fig. 4A. The measured PSFs across the wavelength range from 650 nm to 1000 nm by a x-polarized incident beam for the achromatic flat lens are depicted in Fig. 4B and C, with the details of the optical measurement setup available in the Materials and Methods section and Fig. S5 in the Supplementary Materials. Figure 4D and E further demonstrate the nearly diffraction-limited focusing within its intended achromatic wavelength range. Here the simulated results are calculated through VDIM by convolving the electric filed of the PSF in the xy plane of the achromatic flat lens with the PSF of the microscope imaging system [43, 44].

A high NA single lens with a hyperbolic phase profile is susceptible to strong off-axis aberrations, which restricts the field of view (FOV). Consequently, achieving wide-field imaging is challenging with large NA single lenses, especially those with diameters in the tens of micrometers, such as our achromatic flat lens. To evaluate the wide-field imaging capabilities, we conducted focusing and imaging tests on the achromatic flat lens with an NA of 0.7 and a radius of 30.0 µm. Detailed procedures are available in the Supplementary Materials (see Fig. S7 for more information). Figure 5 displays the microscope imaging results of element 6, group 8 of the United States Air Force (USAF) resolution target under broadband illumination (650 nm – 1000 nm). The achromatic flat lens achieved a resolution of 1.1 µm. During image processing, adjustments were made to the brightness and contrast to enhance image quality[45, 46] (refer to Fig. S8 in the Supplementary Materials for the original images).

## DISCUSSION AND CONCLUSIONS

Table 1 provides a comparison of previously reported experimental achromatic flat lenses in the visible and NIR wavebands for imaging purposes. Traditional broadband achromatic metalenses, which depend solely on the phase dispersion of meta-atoms, face stringent limitations on their NAs and sizes. While achromatic MDLs have the advantage of enabling larger device sizes, they encounter fabrication and efficiency challenges in achieving high NAs. In contrast, it is important to highlight that our achromatic flat lens stands out by effectively balancing both high NA and high imaging resolution, showcasing its competitive edge in the field.

Table1 Summary of previously reported experimental achromatic flat lens in the visible and NIR wavebands for imaging.

| Reference | NA | Microscope Imaging Resolution (μm) | Wavelength range (nm) | Radius (μm) | Maximum Phase dispersion $\Delta\Phi$ | Structure Type |
|---|---|---|---|---|---|---|
| This work | 0.9 / 0.7 | N/A / 1.1 μm | 650 - 1000 | 20.1 / 30.0 | $2.20\pi$ | Meta-atoms on a stepwise substrate |
| Wang et al. [26] | 0.24 | 2.19 μm | 650 - 1000 | 15 | $1.97\pi$ | Metalens |
| Wang et al. [25] | 0.106 | 3.1 μm | 400 - 660 | 25 | $2.62\pi$ | Metalens |
| Hu et al. [19] | 0.164 | 8.77 μm | 420 - 1000 | 25 | $5.70\pi$ | Metalens |
| Chen et al. [18] | 0.2 / 0.02 | N/A / 14 μm | 470 - 670 | 12.86 / 110 | $1.65\pi$ / $1.40\pi$ | Metalens |
| Shrestha et al. [23] | 0.88 | N/A | 1200 - 1400 | 50 | $7.10\pi$ | Metalens |
| Meem et al. [36] | 0.3 | 2.19 μm | 450 - 1000 | 1572.5 | N/A | MDL |

| Xiao et al. [37] | 0.1 | 12.40 μm | 400 - 1100 | 5120 | N/A | MDL |
| Pan et al. [21] | 0.7 | N/A | 400 - 800 | 10 | N/A | Multilayer MDL |
| | 0.5 | 6.20 μm | | 10 | | |
| Richards et al. [22] | 0.3 | 3.91 μm | 400 - 700 | 45 | N/A | Hybird microlens |

In conclusion, by zone division multiplexing of meta-atoms which is achieved by integrating a metalens onto a composite substrate with stepwise thickness, we have successfully created a flat lens that boasts a high NA of 0.9 and a radius of 20.1 μm, demonstrating a broad achromatic bandwidth from 650 nm to 1000 nm in experiment. Compared to a conventional achromatic metalens with the same focal length and a maximum meta-atom phase dispersion of $2.2\pi$, our achromatic flat lens enhances the maximum NA from 0.56 to 0.9 and increases the radius from 6.6 μm to 20.1 μm. Theoretically, conventional achromatic metalenses require a much higher maximum phase dispersion of meta-atoms of at least $13.6\pi$ to deliver similar performance. Furthermore, we fabricated an achromatic flat lens with an NA of 0.7 and a radius of 30.0 μm, covering the same bandwidth, and achieved broadband microscopic imaging with an impressive resolution of 1.1 μm. This resolution surpasses that of both metalenses and MDLs, which do not achieve resolutions finer than 2 μm. The exceptional achromatic imaging capability of these lenses paves the way for their use in high-performance biological microscopy systems, especially in portable and integrated devices. The possibility of creating achromatic flat lenses with even higher NAs and larger diameters is conceivable through the addition of more steps or through optimizing and enhancing meta-atom phase dispersion. Additionally, our methodology for designing broadband achromatic metalenses, along with the fabrication technique, can be adapted to other wavebands, broadening their applicability across various fields and scenarios. This highlights the significant potential and versatility of such lenses in advancing technological applications

## ACKNOWLEDGEMENTS

This work was supported by the National Key R&D Program of China (No. 2021YFA1400800), National Natural Science Foundation of China (Nos. 12374363, 12074444, and 11704421), Guangdong Basic and Applied Basic Research Foundation (No. 2020B0301030009). H.L. acknowledges the "GDTZ" plan support under contract No. 2021TQ06X161. J.Li and X.H.W are supported by Guangdong Provincial Quantum Science Strategic Initiative) (GDZX2306002).

## AUTHOR CONTRIBUTIONS

J.Li. and X.H.W. conceived the idea. J.Lin and J.Li designed and simulated the achromatic flat lens. J.C., J.Lin, and J.Z. fabricated the achromatic flat lens. J.Lin, J.Z., and H.L. performed the achromatic flat lens characterizations and imaging test. All the authors discussed the results. J.Lin, H.L., J.Li, and X.H.W. wrote the manuscript with inputs from all authors. H.L., J.Li. and X.H.W. supervised the project.

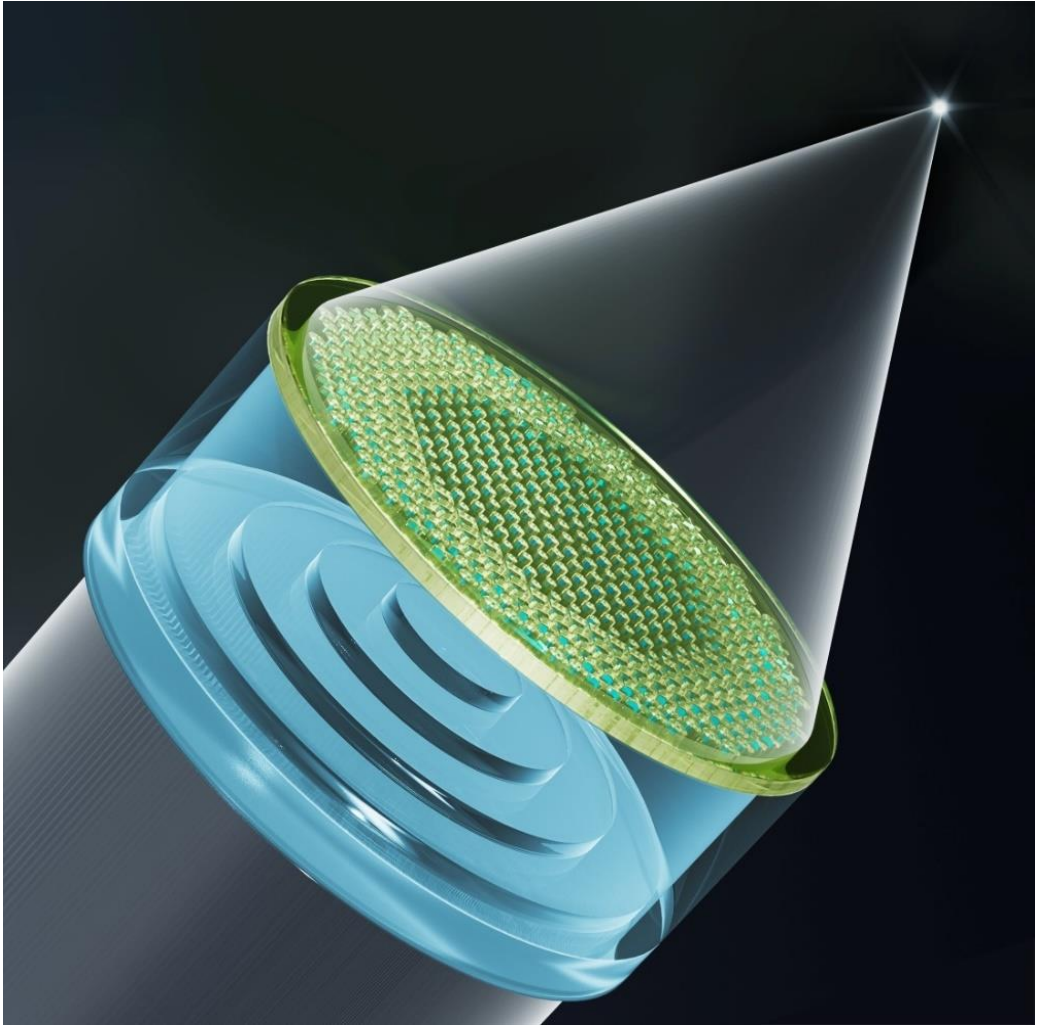

**Fig.1. Schematic of an achromatic flat lens.** An achromatic flat lens consists of a metalens and a stepwise substrate.

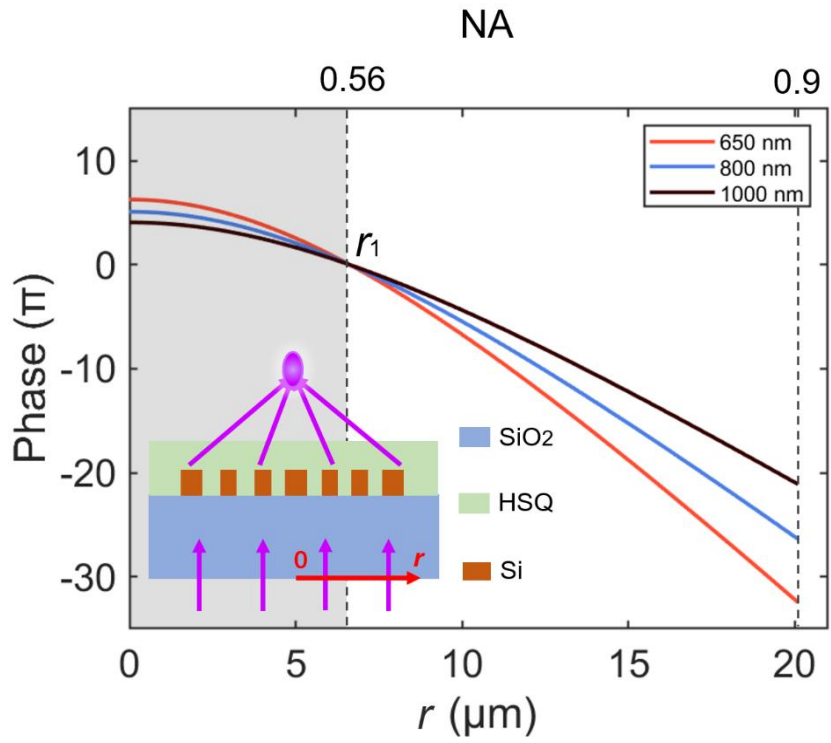

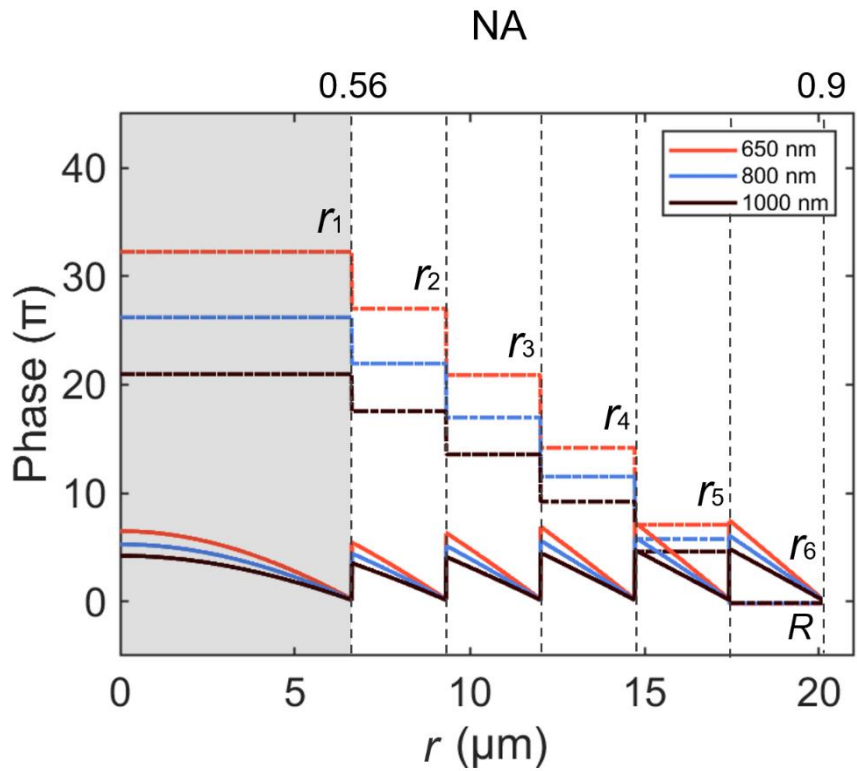

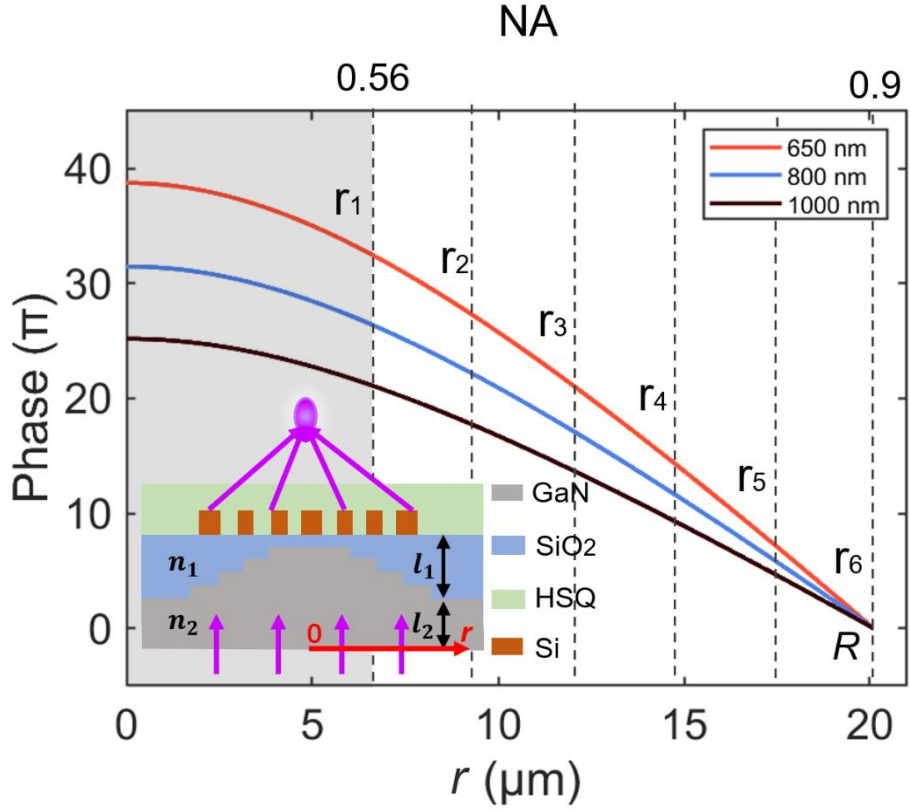

**Fig.2. Design of achromatic flat lens with high NA. (A)** Phase profiles of the conventional achromatic metalens. The focus length $f$ and the maximum phase dispersion of the meta-atoms are set to be 9.7 μm and $2.2\pi$ respectively. The achromatic operating is set within the wavelength range of 650 nm to 1000 nm. The gray region indicates the inherent size limitation of this achromatic flat lens if it is solely composed of conventional metalens. The corresponding maximum radius and NA of the conventional achromatic metalens are $r_0 = 6.6$ μm and NA = 0.56, respectively. Inset is the schematic of the metalens. **(B)** The phase profiles of the meta-atoms $\varphi_{meta}(r, \omega)$ (solid lines), the relative phase $\Delta\omega\Delta L(r)/c$ corresponding to the stepwise substrate (dashed lines) and **(C)** the relative phase $\varphi_{lens}(r,\omega) - \varphi_{lens}(R,\omega)$ of the achromatic flat lens shown in Fig. 1. The parameters are identical to those in Fig. 2A. The corresponding maximum radius and NA of the flat lens are $r_6 = 20.1$ μm and NA = 0.9, respectively. If using a conventional achromatic metalens to achieve the phase control of **(C)**, then the required maximum phase dispersion of the meta-atoms would need to reach $\Delta\Phi_{meta} = 13.6\pi$. Inset is the schematic of the achromatic flat lens.

**A**

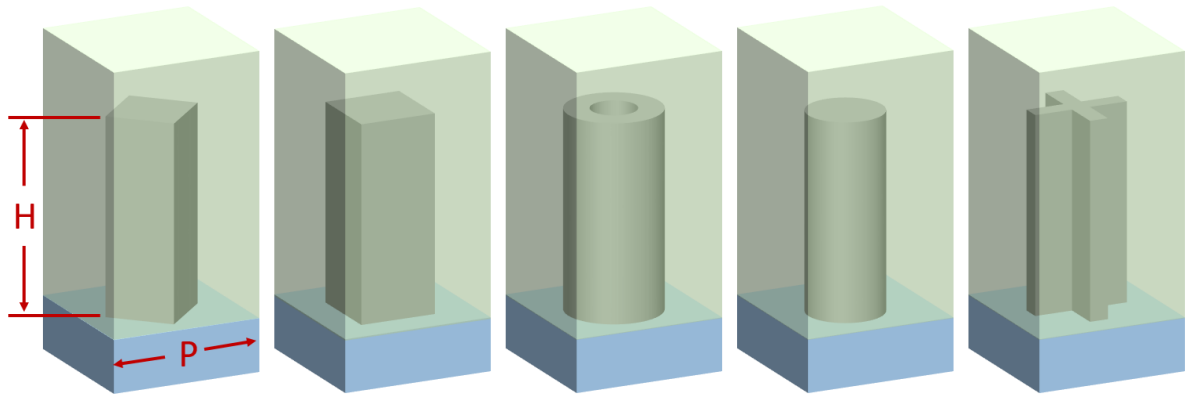

**B**

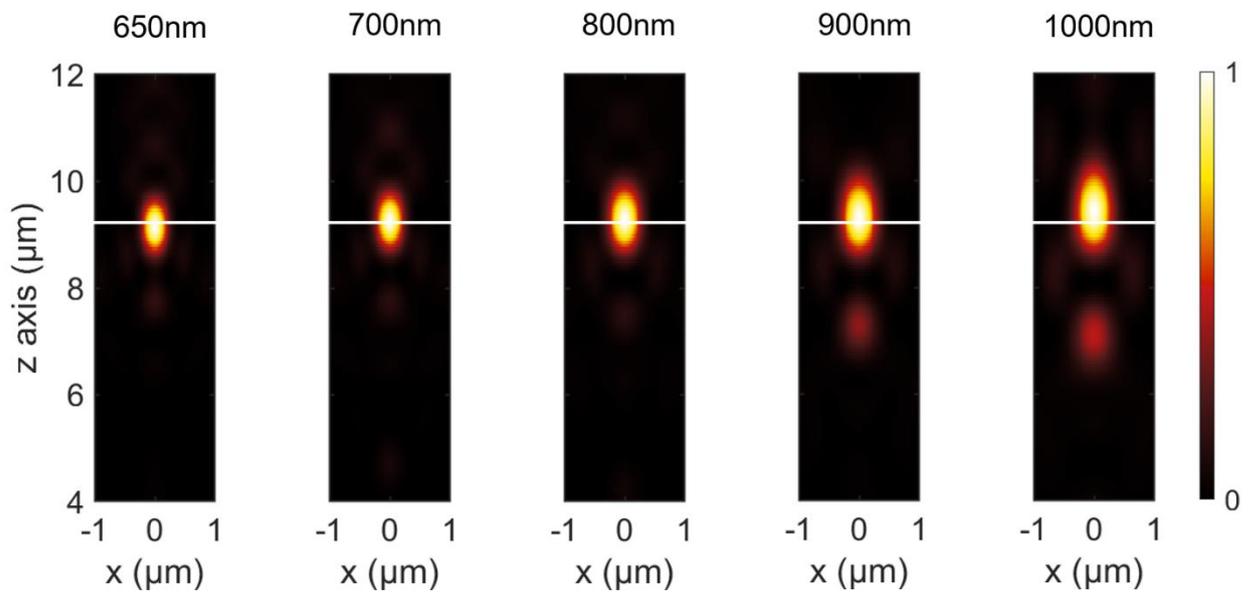

**C**

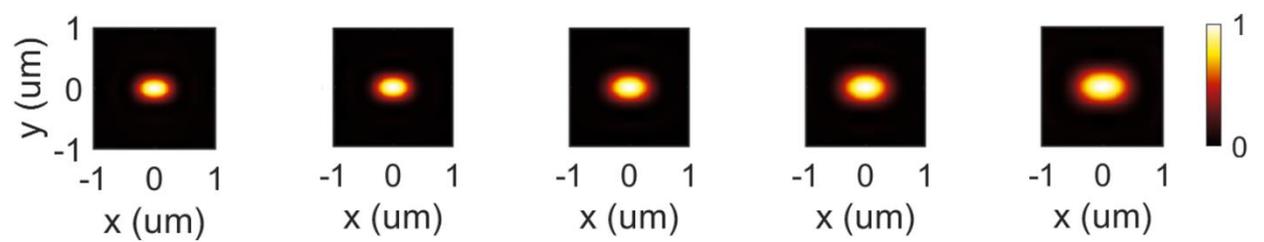

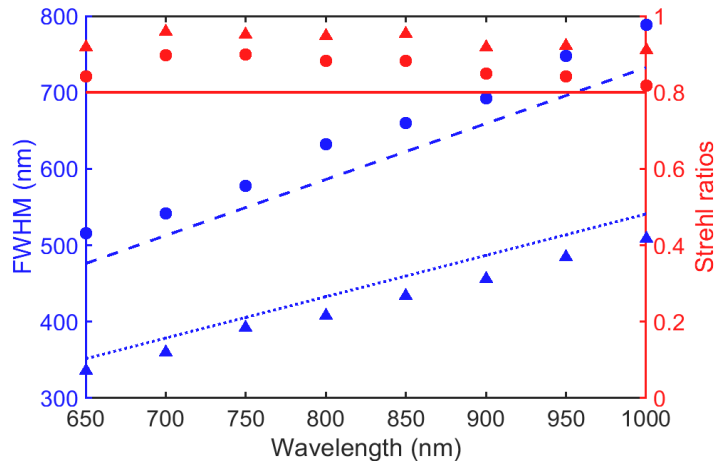

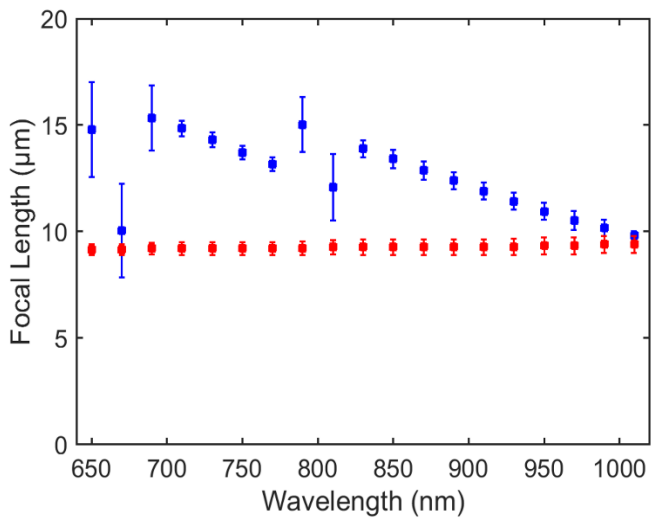

**Fig.3. Simulation of achromatic flat lens with NA = 0.9 and a radius of 20.1 μm.** (**A**) Illustration of five symmetric-shaped meta-atoms composed of c-Si nanostructures. (**B**) The longitudinal PSFs of the achromatic flat lens at different incident wavelength. The white lines indicate the focal plane. (**C**) The corresponding transverse PSFs at the focal plane. (**D**) Calculated FWHM and SR at different incident wavelength for the achromatic flat lens at x direction (round dots) and y direction (triangle dots). The solid, dashed and dotted lines represent the SR above 0.8, theoretical FWHM calculated through VDIM in x direction and y direction, respectively. (**E**) The focal lengths as a function of wavelength for the achromatic flat lens (red dots) and a conventional achromatic metalens (blue dots). The error bars represent the DOF for each focal point.

**A**

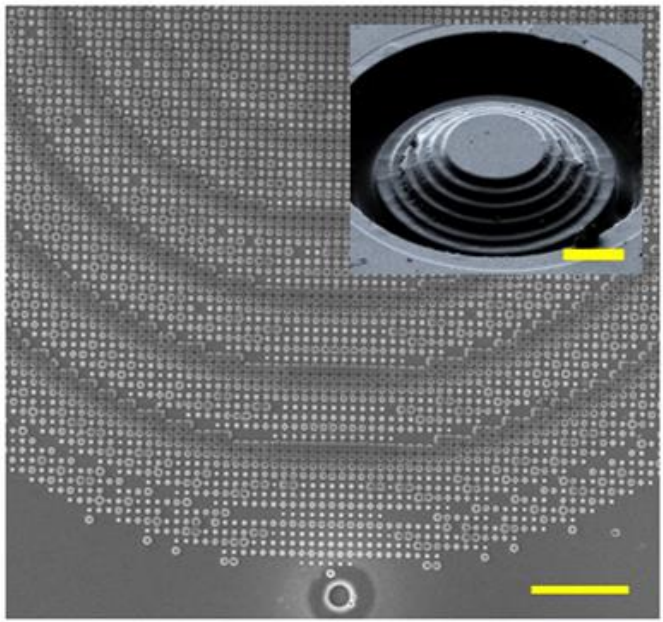

**B**

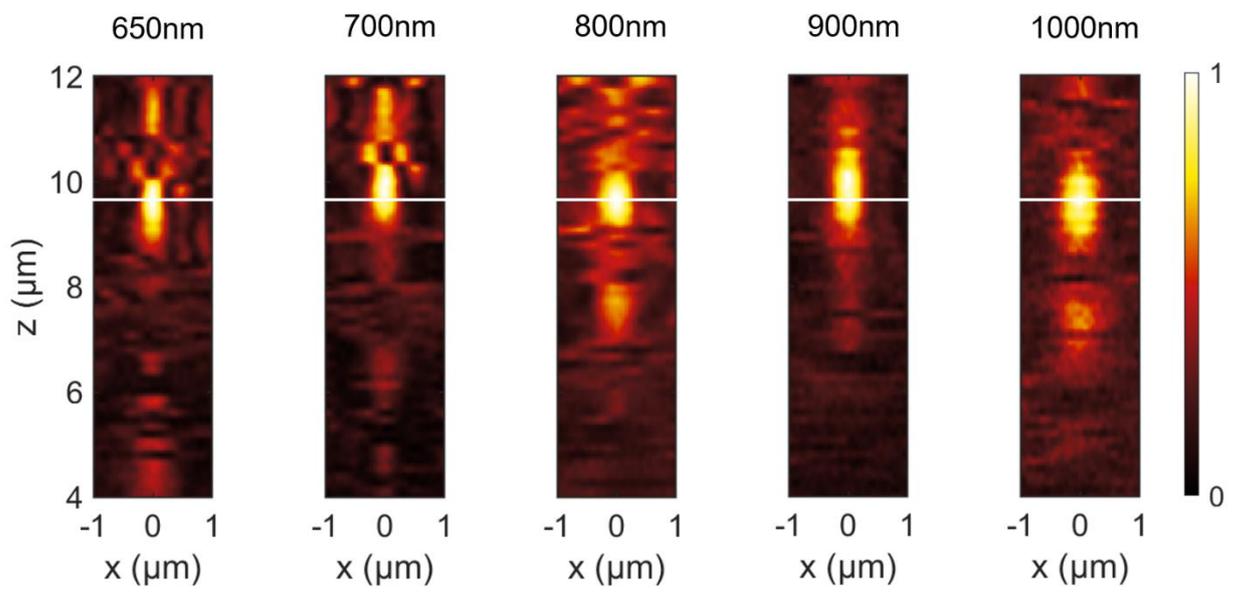

**C**

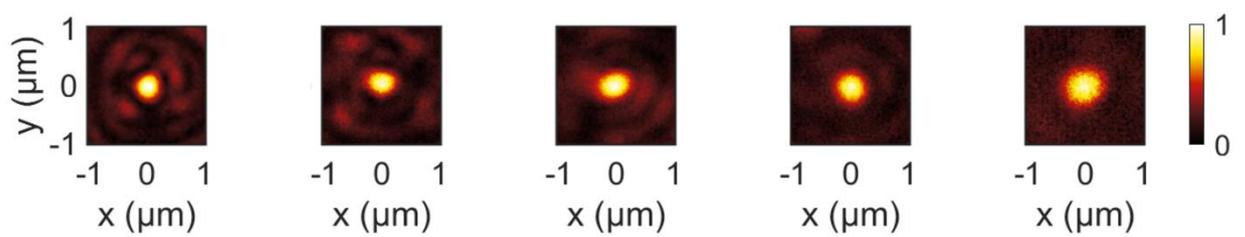

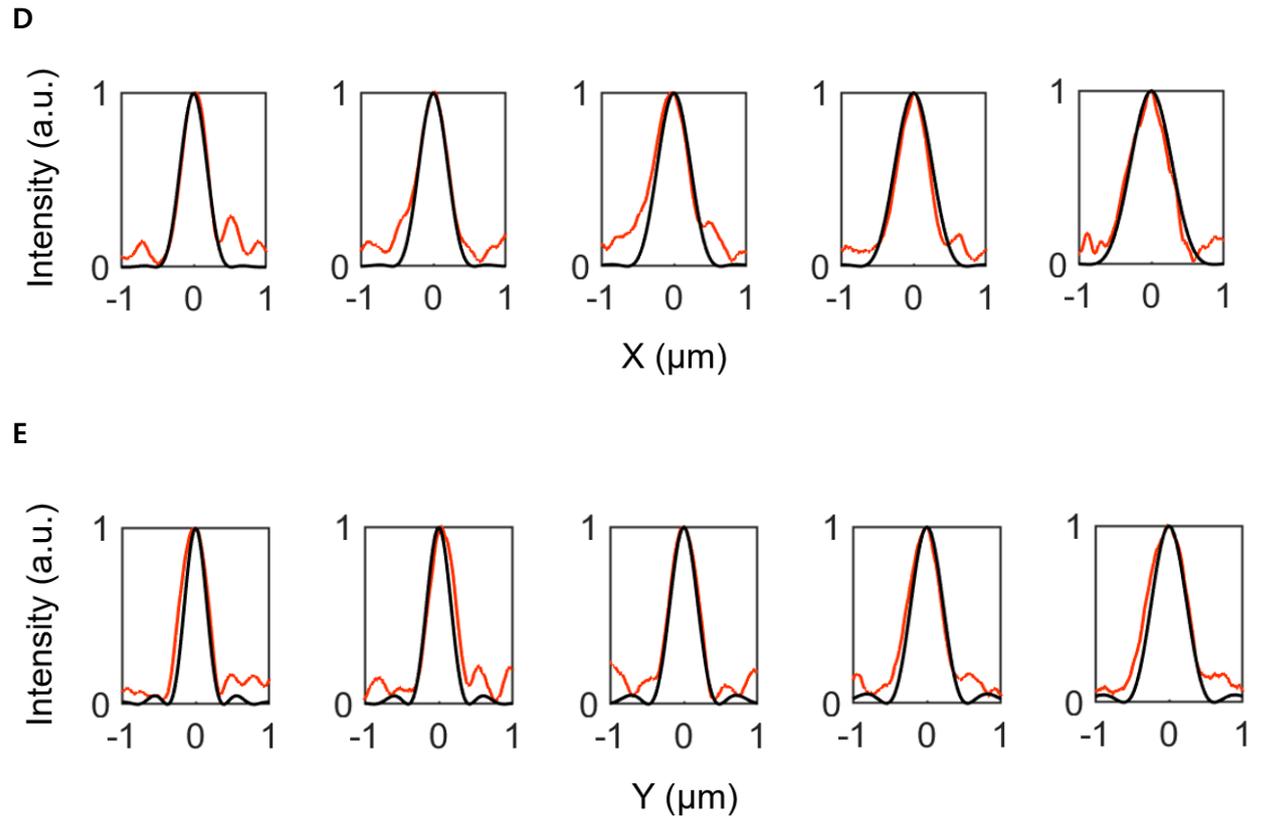

**Fig.4. The experimentally fabricated achromatic flat lens with NA = 0.9 and a radius of 20.1 μm.** (**A**) Scanning electron micrograph (SEM) of the metalens (Scale bar, 3 μm). Inset is the stepwise GaN substrate (Scale bar, 10 μm). (**B**) Experimentally measured longitudinal PSFs at different incident wavelength by a x-polarized incident beam. The white lines indicate the focal plane. (**C**) The corresponding transverse PSFs at the focal plane. (**D**) The normalized intensity profiles at the focus plane in experimental (red lines) and in simulation through VDIM (black lines) at x direction. The FWHMs of the focal spot in experimental are respectively determined to be 401 nm, 458 nm, 495 nm, 524 nm, and 656 nm at wavelengths of 650 nm, 700 nm, 800 nm, 900 nm, and 1000 nm. (**E**) The normalized intensity profiles at the focus plane in experimental (red lines) and in simulation through VDIM (black lines) at y direction. The FWHMs of the focal spot in experimental are respectively determined to be 388

nm, 402 nm, 441nm, 480 nm, 574 nm at wavelengths of 650 nm, 700 nm, 800 nm, 900 nm, and 1000 nm.

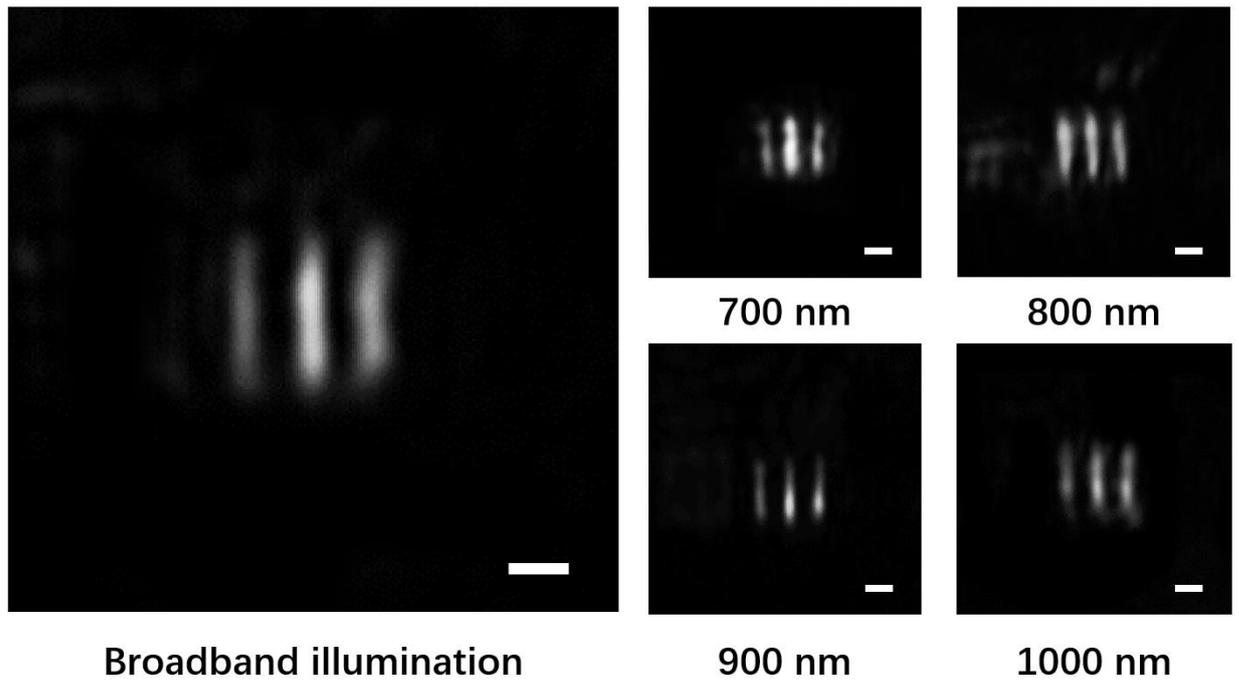

**Fig.5. Imaging using an achromatic flat lens with NA = 0.7 and a radius of 30.0 μm.** The microscope image of element 6, group 8 of the 1951 Unites States Air Force resolution target under broadband, 700 nm, 800 nm, 900 nm and 1000 nm illumination. Scale bars: 2 μm.